\documentclass[sigconf]{acmart}

\AtBeginDocument{%
  }

\setcopyright{acmlicensed}
\copyrightyear{2023}
\acmYear{2023}

\acmConference[SaTS '23] {Proceedings of the 2023 ACM Workshop on Secure and Trustworthy Superapps}{November 26, 2023}{Copenhagen, Denmark.}
\acmBooktitle{Proceedings of the 2023 ACM Workshop on Secure and Trustworthy Superapps (SaTS '23), November 26, 2023, Copenhagen, Denmark}
\acmPrice{15.00}
\acmISBN{979-8-4007-0258-7/23/11}
\acmDOI{10.1145/3605762.3624432}

\begin{document}

\title{Systematic Analysis of Security and Vulnerabilities in Miniapps}

\author{Yuyang Han, Xu Ji, Zhiqiang Wang}
\affiliation{%
  \institution{Beijing Electronic Science and Technology Institute}
  \city{Beijing}
  \country{China}
}

\author{Jianyi Zhang}
\authornote{Corresponding authors}
\affiliation{%
  \institution{Beijing Electronic Science and Technology Institute}
  \city{Beijing}
  \country{China}
}

\renewcommand{\shortauthors}{Yuyang Han, Xu Ji, Zhiqiang Wang \& Jianyi Zhang}

\begin{abstract}
The past few years have witnessed a boom of miniapps, as lightweight applications, miniapps are of great importance in the mobile internet sector. Consequently, the security of miniapps can directly impact compromising the integrity of sensitive data, posing a potential threat to user privacy. However, after a thorough review of the various research efforts in miniapp security, we found that their actions in researching the safety of miniapp web interfaces are limited. This paper proposes a triad threat model focusing on users, servers and attackers to mitigate the security risk of miniapps. By following the principle of least privilege and the direction of permission consistency, we design a novel analysis framework for the security risk assessment of miniapps by this model. Then, we analyzed the correlation between the security risk assessment and the threat model associated with the miniapp. This analysis led to identifying potential scopes and categorisations with security risks.
In the case study, we identify nine major categories of vulnerability issues, such as SQL injection, logical vulnerabilities and cross-site scripting. We also assessed a total of 50,628 security risk hazards and provided specific examples.

\end{abstract}






\maketitle

\section{Introduction}
According to the Internet Society of China, the miniapp has emerged as a new genre of application that has attracted widespread attention in recent years\cite{sgpjbg01}. As of 2019, the total number of miniapp users in China has reached an impressive 1.4 billion, with 780 million new users and 960 million active users. A miniapp is a sub-application that runs within a mobile application, serving as a lightweight and agile solution; that provides users with fast and convenient functions and services. This type of application has gained remarkable traction on various social networking platforms, especially on WeChat, one of the most widely used communication applications with 1.3 billion monthly active users\cite{Zhongguancun03}. Although miniapps are essentially web-based, they provide a seamless and native app-like user experience, allowing users to access various services without installing separate applications. Due to their efficiency and ease of use, these miniapps have gained significant popularity within the mobile Internet ecosystem and have attracted considerable attention from users and businesses.

However, as miniapps continue to flourish, concerns about their security have escalated. As with any web application, ensuring security and protecting user data constitute pivotal aspects that demand careful consideration and proactive measures. This concern has led the official WeChat miniapps to introduce a set of security guidelines\cite{Wechat04}. Nevertheless, during their widespread adoption, miniapps have also revealed vulnerabilities and security issues, attracting the attention of hackers and attackers. Consequently, security has emerged as a pressing issue that needs to be addressed within the miniapps ecosystem. According to the Aladdin Research Institute, the major platforms exert guidance and constraints on miniapps safety through two different models. Firstly, the product model, embodied by WeChat and Alipay, employs an official miniapp security testing service to thwart potential threats proactively. Secondly, the rating model, exemplified by Tiktok, Kuaishou, and Baidu, uses scoring to intervene in security events. In addition, third-party markets offer many security products and services to address and prevent various problems\cite{Aladdin02}.

Meanwhile, security experts have delved into various aspects of miniapps in recent years to protect user privacy and data integrity and to facilitate a thriving miniapp ecosystem. Some have performed backward decompilation of miniapps, while others have examined the differences between miniapps in Android and iOS clients or explored the variances across different platforms. Yet, it is noteworthy that research explicitly targeting the security of miniapp web interfaces and their associated vulnerabilities still needs to be explored.

In this paper, we focus on a comprehensive analysis of the security challenges posed by the web interfaces of miniapps. To this end, we propose a powerful triad threat model for holistically assessing and mitigating security threats. The model strictly adheres to the principle of least privilege\cite{lp} and the direction of privilege consistency. We also propose a new analytical framework for analyzing and evaluating miniapps. This model can identify, diagnose, and mitigate the security risks inherent in miniapps, strengthening the overall security defences and fostering a more secure and trustworthy miniapp ecosystem. Our research has systematically focused on the web interface security of miniapps, particularly emphasising further investigation and exploration of vulnerabilities in miniapps. During the experiment, we identified nine categories of vulnerability issues and 50,628 security risks and threats. We provided examples for each vulnerability issue and recorded an assessment of the security risks.\\
\textbf{Contributions.} We make the following contributions:
\begin{itemize}
\item {We presented a new triad threat model, aiming to gain a more comprehensive insight into the security intricacies of miniapps. Considering the viewpoints of users, the server side, and potential attackers, and under the guidance of least privilege and privilege consistency principles, this model evaluates the security threats and vulnerabilities encountered by miniapps.}
\item {We have constructed a new analysis framework and applied it to traditional vulnerability detection. For the first time, this framework utilizes a unique combination of vulnerability detection, involving AppScan+Proxifier+WeChat(Windows), to detect nearly a thousand well-known mainstream miniapps. This accomplishment represents an essential milestone in miniapp security assessment.}
\item {We conducted a comprehensive evaluation of the novel framework on real-world WeChat miniapps. We meticulously analyzed the detection results, which resulted in identifying nine distinct categories of significant vulnerability issues, along with discovering a total of 50,628 security risks and dangers.}
\end{itemize}

\section{Related work}
Recently, the novel concept of miniapps has been delved into across various domains. Regarding the miniapps, they can be used on education\cite{e1,e2,e3}, online shopping\cite{s1}, healthcare\cite{h1,h2}, campus services\cite{school1},transportation\cite{t1}, food service\cite{f1}. For instance, Zhang et al.\cite{zhang05} devised mini-crawlers to procure miniapps and conducted extensive measurements. Similarly, Zhang et al.\cite{zhang06} explored the realm of identity obfuscation within webview-based super apps, while Liu et al.\cite{liu07} fashioned a dynamic analysis framework tailored for WeChat miniapps. Yang et al.\cite{yang14} also investigated the vulnerability of cross-miniapp redirection among miniapps in WeChat and Baidu. Additionally, Lu et al.\cite{lu08} thoroughly examined the miniapp paradigm, with a specific focus on access control mechanisms. Recently, Wang et al.\cite{wang09} delved into the challenge of concealing APIs in mobile super apps. Meanwhile, Zhang et al.\cite{zhang10} began an inquiry regarding AppSecret leakage in miniapps. Furthermore, Wang et al.\cite{wang11} introduced an innovative approach named TAINTMINI for detecting sensitive data flow in miniapps.

OWASP (Open Web Application Security Project)\cite{OWASP12} and WASC (Web Application Security Consortium)\cite{wasc13} are esteemed organizations devoted to fortifying the security of Web applications. They assume a pivotal role in the assessment of web security. In this research study, we adeptly integrate the identification principles of OWASP Top 10 and the security guidelines provided by WASC, delving into exploring vulnerabilities and security evaluation in miniapp security. Our goal is to assess the security of miniapps systematically, precisely pinpoint potential vulnerabilities, and lay a foundation for further in-depth analysis.

\section{Motivation and Problem Statement}

During the thriving expansion of miniapps embedded within mobile applications, an increasing number of users and enterprises are harnessing these diminutive wonders to enhance their mobile endeavors. Nonetheless, miniapps face a substantial spectrum of security threats and vulnerabilities. Previous research has primarily focused on isolated security issues, resulting in the need for a holistic perspective and revealing gaps in comprehensive analysis concerning the harmonious convergence of users, servers, and potential attackers.

\subsection{Threat Model}
We present a comprehensive triad model for improving the security of miniapp web interfaces and proactively addressing vulnerabilities. This model considers the viewpoints of users, servers, and potential attackers, with the primary goal of providing miniapps with robust resilience to mitigate security threats effectively. The model integrates proactive measures to prevent vulnerabilities with passive strategies to mitigate potential attacks. The methodology used in this model is characterized by its systematic and meticulous nature. It is an enlightening guide for developers, enabling them to assess, improve and maintain miniapps' security.
The triad model can be meticulously designed to strengthen the security of web interfaces and address the vulnerabilities inherent in miniapps. The model has three primary dimensions:
\begin{itemize}
\item {\texttt{Users}}: Representing the clientele of miniapps, users are susceptible to security threats and risks, while their actions can also impact the miniapps overall security posture.
\item{\texttt{Servers}}: Serving as pivotal communication hubs between miniapps and backend servers, the security of these nodes directly governs the safeguarding of user data and system dependability.
\item{\texttt{Attackers}}: Potentially posing security threats, attackers may employ diverse tactics to assail miniapps, pilfer user data, disrupt servers, tamper with data, and more.
\end{itemize}

At the same time, the notion of the "system side" is introduced. Within the triad model, the system side encompasses the system layer, which includes the operating system, the host system, and the miniapp system itself. The dynamic interaction between the security aspect (including the three components of the triad model) and the system side, together with the intricate interplay between the system side, the user side, the server side, and the attacker side, collectively create the comprehensive security ecosystem of the miniapp system.

\subsection{Key Observations}
\label{s3.2}
As shown in Figure 1, we present three potential attack vectors to which the triad model responds.
\begin{figure}[h]
  \centering
  \includegraphics[width=0.7\linewidth]{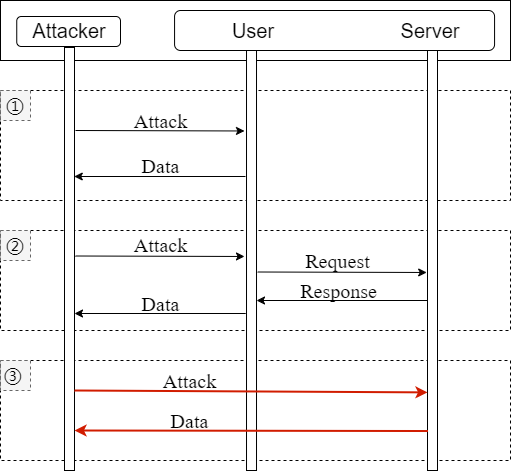}
  \caption{There are three ways of potential attacks.}
  \Description{There are three ways of potential attacks.}
\end{figure}

\textbf{An attacker directly targets the user:}
The attacker endeavors to strike directly at the user, aiming to exploit the user's vulnerabilities or missteps to compromise the user's device or extract sensitive information. This may involve employing phishing techniques, deploying malicious links, introducing malware, and more. 

\textbf{The attacker attacks the server side through the user:}
The attacker adopts an indirect approach to target the server side by manipulating user behavior. By inducing a user to visit a malicious website or download a file containing malevolent code, the attacker manages to infiltrate the server side with the malicious payload. 

\textbf{The attacker directly attacks the server side:}
The attacker launches a direct assault on the server side, striving to take control of it or tamper with server side data. This type of attack may encompass SQL injection\cite{sql}, remote code execution, denial-of-service attacks, and other malicious activities.

Taken together, the triad model provides an integrated security framework that requires attackers to consider the principles of least privilege and privilege consistency, whether they are attacking users directly or exploiting user behavior to target the server side. This approach helps to improve the security of systems and prevent the creation and exploitation of vulnerabilities. In addition, unified attack and defense enable security organizations to gain a deeper understanding of attacker behavior and motivations, enabling more effective mitigation of potential threats.

Meanwhile, each facet of the triad framework assumes a pivotal role in distinct stages of the system side:
\begin{itemize}
\item {\texttt{User-centric security}}: User awareness and authentication mechanisms are crucial in empowering users to safeguard their data and privacy during interactions with miniapps and host systems.
\item{\texttt{Server fortification}}: Host system providers must ensure the secure development and deployment of their applications, including miniapps, to protect user data and maintain the integrity of the system.
\item{\texttt{Attacker mitigation}}: Preventing potential attackers is crucial for the overall system security, safeguarding miniapps and other applications from potential threats.
\end{itemize}

\subsection{Problem Statement and Scope}
The triad model exhibits the following exquisite characteristics:
\begin{itemize}
\item {\texttt{Comprehensiveness}}: The triad model skilfully encompasses the roles of users, servers, and attackers, converging with the system side to provide a comprehensive security research framework. This methodology facilitates the conduct of multifaceted investigations into the security of miniapps, thereby giving deep insights into the intricate security aspects of the system.
\item{\texttt{Systemic}}: Unlike a narrow focus on specific security vulnerabilities, the triad model takes a holistic approach to addressing security concerns throughout the miniapp lifecycle. It proposes more effective security solutions by addressing the root causes of security problems.
\item{\texttt{Extensibility}}: Seamlessly integrated with the system side, the triad model offers remarkable flexibility and scalability. This versatility allows it to be applied to different types and sizes of miniapps while allowing for adaptations and enhancements to address evolving technologies and threats.
\item{\texttt{Responsive to Diversity}}: The variety and complexity of security issues within miniapps makes traditional one-dimensional research methods inadequate. The introduction of the triad model facilitates a comprehensive exploration of miniapp security from multiple dimensions and perspectives, leading to accurate and effective solutions tailored to various scenarios.
\end{itemize}

In prior work related to miniapps, the security issues inherent in these applications can be comprehensively categorized from the perspectives of both attackers and victims.

\begin{itemize}
\item {\texttt{\textbf{Attacker's perspective}}}: From the perspective of potential attackers, their intentions revolve around obtaining sensitive user information, stealing data, disrupting the regular operation of systems, or spreading malicious code. They use methods such as XSS, CSRF, and SQL injection to target the web interfaces of miniapps. This method involves manipulating request parameters, forging requests, and using other mechanisms to carry out their attacks. In addition, they may use techniques such as fuzz testing and penetration testing to uncover vulnerabilities in miniapps, including instances such as unauthorized access and parameter injection, to determine their attack vectors.
\item{\texttt{\textbf{Victim’s perspective}}}: Users could face various risks, ranging from the potential exposure of personal privacy to the risk of account compromise. When using miniapps, users should increase their security vigilance, avoid malicious links and exercise caution when authorizing permissions. Conversely, if servers are compromised, they could face problems such as exposure to sensitive data and system downtime. Therefore, it is vital to strengthen security measures and examine and filter incoming data to prevent attacks.
\end{itemize}

The triad model provides a new perspective and approach to exploring and solving security challenges within miniapps. Within this model, a holistic approach to security is essential, encompassing both offensive and defensive perspectives to protect user information. Following the principle of least privilege, granting users only the rights they need can reduce the risk of breaches and minimize unauthorized operations and data access. At the same time, maintaining privilege consistency ensures consistent user authentication and validation across the miniapp system, preventing vulnerabilities due to inconsistent auditing privileges.

In addition, the security of miniapps web interfaces requires significant consideration, considering offensive and defensive strategies. Our primary goal goes beyond simply thwarting the malicious efforts of potential attackers; it includes proactively identifying and remediating security vulnerabilities. In addition, we recognize that implementing robust interface authentication and authorization mechanisms can effectively prevent malicious users and attackers from directly accessing sensitive data or performing dangerous operations via interfaces.

To ensure the security of user information, developers must adopt secure coding practices and perform comprehensive code audits during the development phase. This proactive approach aims to identify potential logical vulnerabilities, control discrepancies, instances of SQL injection, and cross-site scripting attacks. At the same time, a thorough security assessment is essential when integrating third-party technologies to prevent the introduction of latent security risks.

Using the triad model in conjunction with a holistic strategy that includes offensive and defensive measures, the principles of least privilege, and privilege consistency, a comprehensive improvement in the overarching security of miniapp web interfaces is achieved. This effort not only preserves user privacy and data integrity but also strengthens miniapps ability to withstand adversarial intrusions.

\vspace{-4mm}

\section{Design}

This section provides the detailed design of the analysis framework, and applies it to traditional vulnerability detection. As described in \ref{s3.2}, security risks in miniapps occur across three components (i.e., users, servers and Attackers) at various granularity (i.e., event handlers, miniapp pages, miniapp programs) in different manners. To address these challenges, the framework uses a novel combination of vulnerability detection with the pairing of AppScan+Proxifier+WeChat(Windows) to detect mainstream miniapps. The triad threat model guides the design and deployment of detection frameworks. By providing a comprehensive understanding of potential threats and vulnerabilities from the user, server and attacker perspectives, the triad model can increase the effectiveness and comprehensiveness of the detection framework. In particular, for the first time, we employed such a combination of detection methods. We used the Windows version of the WeChat client to access the selection miniapp and subsequently directed the data traffic to AppScan through a Proxifier proxy for the purpose of security risk detection.
The structure of this frame is shown in Figure 2:
\begin{figure}[]
  \centering
  \includegraphics[width=0.9\linewidth]{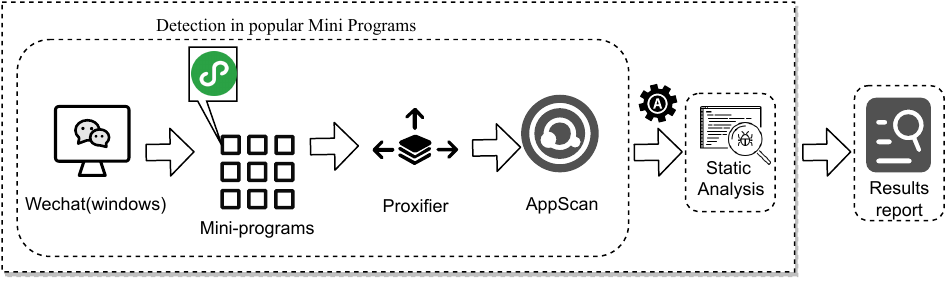}
  \vspace{-3mm}
  \caption{Overview of Analyze Framework}
\end{figure}

\textbf{Scope.}
In this study, we focus on systematically studying the security of miniapps to understand their root causes, and consequences. Given the security concerns around web interfaces, our goal is to identify vulnerabilities within miniapps that arise from web interface complications. Given WeChat's prominent position as a primary social media platform, coupled with the rapid expansion of its miniapp ecosystem, this research assumes notable importance. Therefore, in this paper, we focus on WeChat miniapp in particular due to its popularity and support for both sensitive data access.

\textbf{Methodology.}
As part of the use of the new framework, we are jointly deploying the following applications in the first instance:
\begin{itemize}
\item {\texttt{WeChat platform for Windows}}: In the previous research focused on miniapps, the primary research platform was mainly focused on the Android ecosystem. The primary research methodology included reverse engineering and decompilation techniques.
\item{\texttt{Proxifier}}: This widely used proxy tool allows applications to connect to the Internet through intermediary servers, facilitating the mediation and management of network traffic.
\item{\texttt{Appscan}}: A robust application security scanning tool, Appscan helps to analyze traffic data relayed through proxies. It identifies critical security risks, performs categorization assessments, and assists with remediation. Appscan takes on the task of dissecting the traffic that passes through the Proxifier proxy and provides a comprehensive evaluation of the underlying security concerns.
\item{\texttt{Burp/Fiddler}}: While capturing packets for WeChat miniapps, these tools can intercept network requests sent and received by the miniapps. They allow systematic analysis of the information encapsulated in these requests.
\item{\texttt{Airtest}}: It, a cross-platform mobile automation testing tool, is used to automate tests and evaluate the performance of mobile applications. It is well suited for automating tests and evaluating performance aspects of mobile applications.
\end{itemize}

Airtest makes it easy to simulate user actions and data retrieval within WeChat miniapps. When testing WeChat miniapps: Airtest helps us to efficiently simulate user interactions and retrieve data from within WeChat miniapps. When trying WeChat miniapps, Airtest can be used to simulate user actions and, by creating test scripts, locate and retrieve items within the miniapp. While simulating user actions, we set up a Proxifier proxy to route the traffic data of the WeChat miniapp through Proxifier to AppScan. Using AppScan, we analyze the traffic data passed through the Proxifier proxy. Next, the test application generates reports and automatically compiles security risks from the documentation. Finally, we perform a static analysis of various security risks to consolidate the data further.

%%新加--
AppScan meticulously examines the miniapp's source code and binary files during this process, particularly during static analysis. It identifies a range of vulnerabilities, including but not limited to SQL injection and cross-site scripting attacks. It also performs a comprehensive dependency assessment, including third-party libraries and components, to identify potential vulnerabilities or security issues. At the same time, AppScan can trace the intricate paths of data flow within miniapps, facilitating early detection of potential data leakage and identifying insecure data processing techniques.

In prior research, BurpSuite, Android WeChat, Emulator, and Fiddler were amalgamated to perform detection tasks. As the WeChat version has been updated, the effectiveness of the former detection method for miniapp on the Android platform and within simulators has gradually decreased. Therefore, we introduce a novel approach, employing a distinct pairing scheme involving Windows WeChat, Proxifier, AppScan, and BurpSuite. Through this setup, we analyze the data traffic information of Windows WeChat miniapps using AppScan, detecting potential vulnerabilities. The results are subsequently cross-verified and validated through BurpSuite and other indispensable tools during the testing phase.

\section{Evaluation}

This section focuses on examining web interface security vulnerabilities within miniapps. These vulnerabilities are methodically grouped into distinct focus areas, covering both the frontend and backend. Among the many vulnerabilities identified are logical inconsistencies, permission management anomalies, SQL injection susceptibility, cross-site scripting infiltration, arbitrary file uploads, compromised password security, and other intricate security nuances. Each classification is accompanied by specific examples that meticulously illustrate the authentic security challenges faced by miniapps.

\subsection{Quantifying the Vulnerabilities Risks}
Within this section, these vulnerabilities are systematically categorised into frontend, backend and other relevant dimensions and presented sequentially for comprehensive analysis.

\subsubsection{Frontend vulnerabilities}
\textbf{\texttt{Logic Vulnerabilities}}: This vulnerability refers to the presence of defects or errors in the logical flow of an application, resulting in behavior that deviates from expected results or presents potential security risks. Typically, these vulnerabilities do not involve syntax errors within the code, but are related to complications in the program's business logic or permission controls. Because of their ability to bypass conventional security measures and exploit input validation errors to remain undetected, they manipulate the standard logical flow of a miniapp. As a result, they can cause insecure permissions, access control issues, unauthorised data retrieval and business logic errors, among other problems.

\textbf{\texttt{Cross-Site Scripting (XSS)}}: This vulnerability attack involves surreptitiously inserting malicious script code into the input fields of miniapp, forcing it to execute within the confines of the user's web browser. It allows malicious actors to perform harmful actions during a user's session, including stealing login credentials, manipulating web page content, and redirecting to malicious online domains.

\subsubsection{Backend vulnerabilities}
\textbf{\texttt{Privilege Escalation}}: Inadvertent misconfigurations of permissions can give unauthorised entities access to resources that should be restricted. More dangerously, these misconfigurations could allow such entities to manipulate critical system components. Specific fields such as "id", "uid", and "UserName", in conjunction with their associated counterparts, require scrutiny. Because they can be used as pathways for parameter traversal, putting sensitive information at significant risk.

\textbf{\texttt{SQL Injection}}: Insufficient user input validation can lead to malicious SQL query injection to access and modify database data. The SQL injection vulnerability in miniapps means that user input is not fully validated and filtered in the backend code of the miniapp. As a result, an attacker can construct malicious inputs and execute malicious SQL statements to gain unauthorized access to the database or to modify, delete or leak data from the database.

\textbf{\texttt{Arbitrary file uploads}}: Owing to the absence of robust file type validation, malicious actors gain the capability to upload pernicious files that pose a grave threat to the security of the system.

\textbf{\texttt{Weak password security}}: Within the miniapp, the vulnerability of weak password security resides in the login authentication or backend management system. Should users choose passwords that are overly simplistic, readily guessable, or commonly employed, the security of their accounts diminishes, rendering them susceptible to password guessing or brute-force cracking attacks. Furthermore, weak passwords may also engender password-guessing attacks, brute-force cracking attempts, and instances of multi-account hijacking.

\subsubsection{Other Security Issues}
\textbf{\texttt{Leakage of sensitive information}}: Inadequate management of sensitive information within miniapps during their design, development or operational phases can create scenarios where malicious entities can inappropriately access or disclose user-sensitive data. Furthermore, in the miniapps, improper handling of user input data within URLs can also expose sensitive information or facilitate unauthorised access to resources.

\textbf{\texttt{Cross-Site Request Forgery}}: This vulnerability presents a scenario where malicious actors impersonate authentic users and trick them into unknowingly executing malicious requests through their web browsers. This manipulation results in the initiation of unauthorised actions within the miniapp. By taking advantage of the user's authenticated session, this vulnerability exploits the limited ability of miniapps to thoroughly validate the source of incoming requests, allowing malicious operations to be covertly executed.

\textbf{\texttt{Integration of Third-party Technologies}}: During the development phase of miniapps, the inclusion of third-party technology components, either developed by external vendors or by fellow developers (such as plug-ins, libraries, APIs, etc.), can create security vulnerabilities and potential threats. These include security breaches, data leakage and the propagation of malicious code.

\subsection{Case Studies}

\subsubsection{Frontend vulnerabilities}
\textbf{Logic Vulnerabilities:} Throughout our experiments, we discovered that a substantial proportion of the vulnerabilities inherent in the miniapp were attributed to logic flaws. Among them, we encountered a category known as "authentication type logic vulnerabilities." These vulnerabilities encompassed a range of security concerns, including login bypass (enabling arbitrary user access), password recovery vulnerabilities, CAPTCHA circumvention, exploitation of mobile phone CAPTCHAs, SMS-based attacks (commonly referred to as SMS bombs), and the masking of mobile phone numbers during registration. 

During the experiment, we encountered several noteworthy logical anomalies, which warrant careful attention:

\begin{itemize}
\item {In our examination of the "Take goods mall" miniapp, we observed that after the login authorization, the miniapp omitted the secondary verification of the user's mobile phone number to ensure data protection. This oversight resulted in direct access to the miniapp without verifying the mobile phone number.}
\item {In our evaluation of the miniapp for the "Haidian Foreign Language Students Service Platform," we input "123," followed by a single quote during the login phase and provided a random password. By capturing the package using tools like BurpSuite, we noticed incorrect account information was returned, yet we could still directly access the system.}
\item {During our assessment of the "Yixianjia Jiadele Life Supermarket Online Store" miniapp, we discovered that entering the verification code "111111" during the registration phase resulted in immediate successful registration without stringent validation of the bound mobile phone number.}
\item {In evaluating the "To Shoot | Milk Delivery Supermarket Recycling Service" miniapp, we observed a lack of robust validation for the verification code during the login phase, allowing unrestricted access using any four-digit code.}
\item {While examining the "Panqinkanghua" miniapp, we attempted to replay the verification code package. We observed multiple SMS messages received on the phone quickly, indicating a vulnerability to SMS bombing.}
\item {Furthermore, while testing the "Yonghui Enterprise Purchase" miniapp, we encountered a four-digit verification code with a five-minute validity period. And upon investigation, we found that the verification code could be traversed to obtain the correct sequence.}
\end{itemize}

\textbf{Cross-Site Scripting (XSS):} In the miniapps "Mom.com Mom's Good Products," "Gourmet Jie Recipe Book," and "Jurong Convenience Supermarket Preferred Life Service Platform," we have strategically inserted prompt messages into the input box, comment section, search box, or any other text fields that users can interact with on the webpage. Our approach strictly adheres to benevolent experimental testing and forbids using malevolent XSS-constructed statements. The attack is deemed valid once the prompt messages can be successfully executed. In the trio mentioned above of miniapps, the prompt statement was entered into the search field, and subsequently, "200 OK" was obtained in BurpSuite.

\subsubsection{Backend vulnerabilities}
\textbf{\texttt{Privilege Escalation}}: An authentication vulnerability through parameters in a cookie can be exploited by manipulating the UserId value to gain unauthorized access to user information. When the UserId value is altered to a different value, the system divulges various essential details about the employee, including their employee number, name, mobile phone number, and work status (either "in" or "out" of work). For instance, in the case of Shangbiao Brand Supermarket, by employing Burp Suite to modify the id-data, the corresponding information for different ids can be accessed, thereby revealing mobile phone numbers, order numbers, and other sensitive information.

\textbf{\texttt{SQL Injection}}: SQL Injection is a prevalent security vulnerability observed in miniapps. Its occurrence can largely be attributed to the inadequate handling of user input data in the backend code, which results in the direct concatenation of user input into SQL query statements. As a consequence, attackers can manipulate the execution logic of the SQL statement by modifying the input, thus gaining unauthorized access to sensitive information and undermining the integrity of the database.
\begin{itemize}
\item {During the examination of the "Jiale Source Fresh Supermarket Corner Store" miniapp, we encountered an SQL injection vulnerability. By skillfully constructing and concatenating SQL query statements in the keyword section, we could access sensitive information from the system, including database details.}

\item {In the context of the "Enterprise Management Cloud" miniapp, we identified an SQL injection vulnerability in the login section. By injecting manipulated SQL statements into the account section, we were able to trigger error messages or, at times, inadvertently disclose sensitive information in pop-up windows.}

\item {Similarly, the "Gourmet Jie Recipe Collection" miniapp displayed SQL injection vulnerabilities in our examinations. We confirmed this by inputting specific values into the id section of the SQL query statement.}

\item {Lastly, multiple SQL injections were detected during testing in the "Water Sage Technology Campus Direct Drinking Water" miniapp. For instance, manipulating the uid section allowed access to sensitive information. Should this miniapp, named "Campus to provide direct drinking water," suffer from information leakage, it could lead to severe consequences.}
\end{itemize}

The Figure 3 shows how an attacker exploits a SQL injection vulnerability to gain unauthorised access to sensitive information.
\begin{figure}[h]
  \centering
  \includegraphics[width=0.9\linewidth]{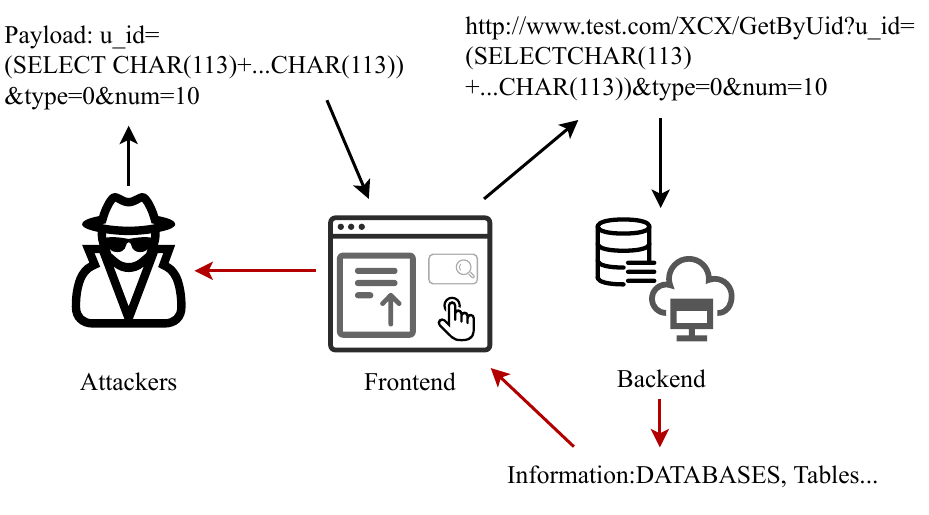}
  \vspace{-3mm}
  \caption{An example of the SQL injection process}
\end{figure}

\textbf{\texttt{Arbitrary file uploads}}: The absence of file type validation permits malevolent files to be uploaded, thus posing a grave threat to system security. Primarily, a burp manipulation of the file extension can exploit any location where an image is uploaded. Secondly, in cases where an ID card is uploaded, if it is done through a camera, the vulnerability can be circumvented by modifying the state in the return packet in conjunction with process validation. For instance, a print miniapp can evade validation and upload an image by altering the file extension via BurpSuite.

\textbf{\texttt{Weak password security}}: During the testing process, we shall come across passwords such as "123456," "111111," "888888," "abc123," and other feeble, simplistic account credentials. This situation may arise due to the miniapps' convenient utilization by employees during the testing phase; however, it inevitably gives rise to security vulnerabilities.

\subsubsection{Other Security Issues}
\textbf{\texttt{Leakage of sensitive information}}: In the decompiled source code of miniapps, user information is susceptible to leakage. Even in some miniapps where source code maintenance is infrequent or has ceased, highly sensitive information, such as uid and key, might be exposed. Furthermore, in certain registration and login interfaces, if incorrect or intentional SQL statements are input consecutively, a prompt box will provide feedback based on the input, allowing for the extraction of sensitive information through subsequent statement crafting. During experimentation, we encountered a miniapp that, despite displaying a prompt box indicating discontinued maintenance, still allowed access to numerous pages from which a wealth of private information could be gleaned through decompilation.

\textbf{\texttt{Cross-Site Request Forgery}}: Insufficient CSRF protection could allow an attacker to execute actions on behalf of an authenticated user surreptitiously. Once the user logs in and saves their credentials within the miniapp, the authorized information remains accessible for subsequent logins. Subsequently, the attacker can employ a malicious link to deceive the user into clicking on it. Due to the absence of an effective CSRF defense mechanism in miniapps, the attacker can exploit the user's active login session to initiate malevolent requests, all unbeknownst to the user.

\textbf{\texttt{Integration of Third-party Technologies}}: During the meticulous testing phase, we encountered many miniapps, totaling more than a dozen, which exhibited a common security flaw -adopting the easily-guessable verification code "11111." Astonishingly, successful registration and login were achieved simply by inputting this rudimentary code. Remarkably, these miniapps were supported by the third-party technological expertise of KM Technology. Regrettably, KM Technology's services vulnerabilities could potentially jeopardize the security and integrity of multiple miniapps affiliated with the company.

\subsection{Evaluation Results}
After eliminating extraneous factors, the conclusions of the scan are as Table~\ref{tab:commands}. Total number of severity issues included in the report: 50,628.

\begin{table*}
  \caption{Security Vulnerabilities in miniapps}  \vspace{-2mm}
  \label{tab:commands}
  \begin{tabular}{ccl}
    \toprule
    Severity Level &A Number & Comments\\
    \midrule
    \texttt{High} & 5,456 & Serious, urgent, need to be fixed immediately.\\
    \texttt{Moderate}& 4,042 & General, higher risk, timely attention.\\
    \texttt{Low}& 35,543& Slight, small risk, timely attention.\\
    \texttt{Informational}& 5,587& Recommendations, tips, can be optimized.\\
    \bottomrule
  \end{tabular}
\end{table*}

In miniapps, application data encompasses cookies, JS, parameters, comments, visited URLs, failed requests, and filtered URLs. Amongst this trove of data, a myriad of security vulnerabilities lies in wait, constituting a profound menace to user data's sanctity and system resources' stability. These vulnerabilities can be systematically categorized into risk levels, ranging from high to medium to low.

\begin{itemize}
\item {High-risk vulnerabilities pose grave repercussions, encompassing the peril of breaching the sanctity of administrative privileges, compromising databases, and executing remote commands on web servers. Moreover, they can potentially unleash debilitating denial-of-service onslaughts, lay bare delicate information like source code and user credentials, and even facilitate illicit transactions.}

\item{Medium-risk vulnerabilities encompass the threat of session hijacking and manipulation, unauthorized access to sensitive resources, and the exposure of confidential data transmitted during encryption. In addition, attackers can subvert authentication mechanisms, infiltrate specific directories, and extract files or configurations from web servers.}

\item{Low-risk vulnerabilities entail gathering sensitive information, luring users into divulging crucial data, and extracting server side script source code. Furthermore, these vulnerabilities may result in information disclosure, downloading of transient script files, and manipulation of client side sessions or cookies to impersonate genuine users.}
\end{itemize}

As is shown in Figure 4, the presented data delineates the distribution of vulnerabilities acquired through framework-based detection, alongside the proportion of vulnerabilities that underwent our comprehensive classification and subsequent meticulous verification encompassing nine prominent vulnerability categories.
\begin{figure}[h]
  \centering
  \includegraphics[width=\linewidth]{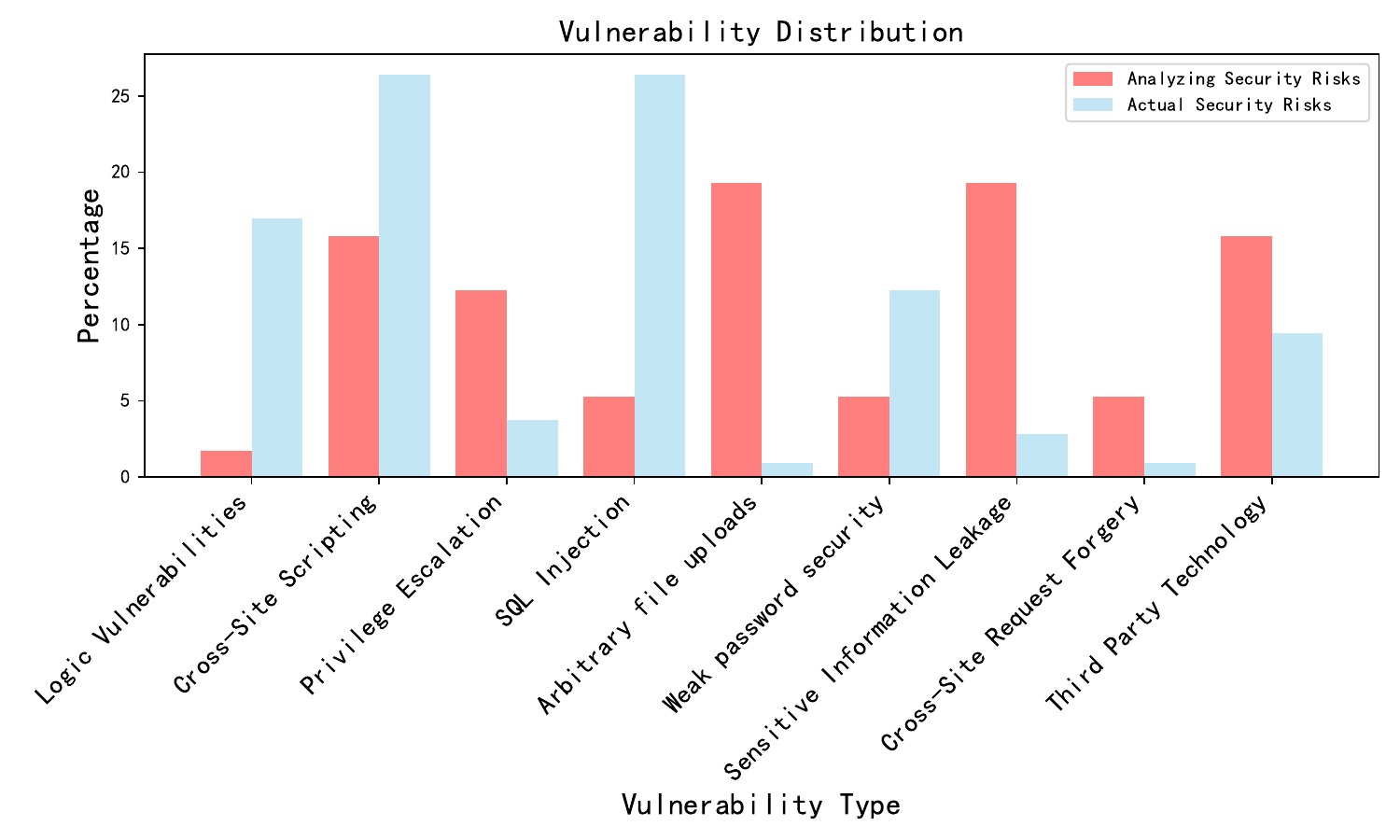}
    \vspace{-7mm}
  \caption{These are nine broad categories of vulnerabilities}
\end{figure}

The data presented in Figure 4 clearly shows that real security threats such as logical vulnerabilities, cross-site scripting and SQL injection manifest themselves at a much higher frequency than their proportional representation in the security risk analysis. This observation highlights the profound dangers of logical vulnerabilities, XSS and SQL injection to the security of miniapps. At the same time, while injection vulnerabilities dominate the landscape of security risk analysis, our study focuses primarily on the tangible instances of SQL injection vulnerabilities, resulting in their comparatively lower representation.

Conversely, the incidence of privilege escalation, sensitive information leakage and arbitrary file upload vulnerabilities in the actual security risk landscape are significantly lower than their representation in the security risk analysis. This disparity is because these facets are predominantly examined through the lens of code audits. This method is separate from our primary approach within this paper.\vspace{-2mm}

\section{Discussion}
The manuscript above delves into an all-encompassing appraisal of the miniapp, focusing on the esteemed OWASP Top 10 and WESC. It diligently explores the security aspect of the miniapps web interface. It conducts thorough research on vulnerability mining, culminating in the proposition of a triad model that fosters a more systematic and comprehensive approach to future investigations. Furthermore, the paper unveils the vulnerabilities unearthed during the experimentation phase. In light of these security concerns, it is imperative to persistently enhance and fortify the following aspects of miniapp security in the times ahead:

\begin{itemize}
\item {\texttt{Interface Security}}: Ensuring the integrity of interfaces is essential to enable seamless communication between miniapps and backend services. These interfaces play a crucial role in the miniapp domain, especially in transferring and managing sensitive user data. Adopting appropriate authentication and authorisation mechanisms, such as \textit{OAuth} or token-based authentication, is paramount in maintaining the security of interfaces.  In addition, implementing rigorous input validation and data filtering is a fundamental safeguard against malicious intrusion, including threats such as SQL injection and XSS attacks.

\item{\texttt{Platform Security}}: Platform security, on the other hand, encompasses creating a safe and sheltered ecosystem for miniapps to thrive within. The safe encompasses safeguarding the miniapp distribution platform with an unwavering commitment to data protection. It is imperative to elevate the significance of passwords that correspond to miniapps and diligently update them promptly. Additionally, developing robust user login processes and establishing periodic security reviews and vulnerability remedies serve as steadfast guardians, ensuring users remain shielded from malicious apps and data breaches.

\item{\texttt{Backend Security}}: Safeguarding the sanctity of the miniapps backend server stands as an imperative in the realm of backend security. Within this domain, the backend assumes responsibility for storing and processing invaluable user data, hence necessitating rigorous access control and robust data encryption measures. Furthermore, the prohibition of external network access to the backend management platform remains a steadfast fortification, while the timely updating of plug-ins, APIs, and other components bolsters defense walls. Regularly tending to server side scripts and databases by applying updates and patches mitigates the risks posed by potential attacks.

\item{\texttt{Staff Responsibilities}}: As the custodians of miniapp development and maintenance, staff have a significant responsibility in ensuring the security of these miniapps. They must undergo comprehensive security training to provide a thorough understanding of security threats and mitigation methods. Adherence to established best practices is a guiding principle in conducting meticulous security assessments that include source code, interfaces, platforms and backend systems. This vigilant approach plays a vital role in the timely identification and remediation of latent security issues, thereby safeguarding the integrity of the miniapps they manage.
\end{itemize}

A cohesive defense can be strengthened by carefully managing factors such as source code security, interface security, platform security, back-end security, and personnel responsibilities, enabling miniapps to deftly fend off potential security threats and maintain user privacy and data integrity. As we diligently address the security challenges inherent in miniapps, future research could address advanced vulnerability detection techniques, the fusion of artificial intelligence and security protocols, studies of the miniapp ecosystem, vulnerability mitigation and management strategies, security assessment, and authentication mechanisms. These pursuits will foster collaborative efforts. Through the relentless pursuit of knowledge and in-depth research, we will continuously improve the security of miniapps, ensure the protection of user information, and promote the sustainable development of the miniapp industry.

\section{Conclusion}

In this paper, we delve into the realm of miniapp security, with a central focus on discerning and mitigating diverse vulnerabilities that have the potential to jeopardize the sanctity of user data and the integrity of the system. To this end, we propose an innovative triad threat model incorporating users, servers, and potential attackers, upholding the tenets of least privilege and privilege consistency to safeguard user information. This model accentuates the significance of comprehensively considering the interplay between these three entities, thereby creatively designing a novel security analysis framework for miniapps. Furthermore, our exploration encompasses an extensive survey and illustrative examples of the manifold security risks looming in miniapps front-end and back-end spheres. By diligently categorizing these issues and furnishing specific instances, we illuminate distinct facets of miniapp security and underscore the criticality of addressing these vulnerabilities. In culmination, we focus on miniapp security and chart a course for future research in this domain.

\section*{Ethics and Responsible Disclosure}

In adherence to established community practices, we employ our accounts and computational resources within a controlled environment for thorough analysis and experimentation aimed at identifying potential security risks. Throughout our investigative process, we dutifully reported the outcomes of the vulnerability assessment to the vendor, offering remediation recommendations, and subsequently obtained acknowledgment of the identified vulnerabilities. Upholding a commitment to strict confidentiality, we safeguard all information details before the vendor addresses and provides feedback on the vulnerabilities, thereby preventing harm to users, miniapp developers, and platform providers.

\begin{acks}
Supported by the Fundamental Research Funds for Central Universities (328202204).
\end{acks}

\bibliographystyle{ACM-Reference-Format}
\balance
\bibliography{sample-base}


\end{document}